\documentclass[journal]{IEEEtran}
\usepackage{amsmath,amsfonts}
\usepackage{algorithmic}
\usepackage{algorithm}
\usepackage{array}
\usepackage[caption=false,font=normalsize,labelfont=sf,textfont=sf]{subfig}
\usepackage{textcomp}
\usepackage{stfloats}
\usepackage{url}
\usepackage{verbatim}
\usepackage{graphicx}
\usepackage{cite}
\begin{document}

\title{Block-QAOA-Aware Detection with Parameter Transfer for Large-Scale MIMO}

\author{Shuai~Zeng%
\thanks{Shuai Zeng is a faculty member with Chongqing University of Posts and Telecommunications, Chongqing 400065, China (e-mail: \mbox{zengshuai@cqupt.edu.cn}).}}

\maketitle

\begin{abstract}
Large-scale MIMO detection remains challenging because exact or near-maximum-likelihood search is difficult to scale, while available quantum resources are insufficient for directly solving full-size detection instances by QAOA. This paper therefore proposes a Block-QAOA-Aware MIMO Detector (BQA-MD), whose primary purpose is to reorganize the detection chain so that it becomes compatible with limited-qubit local quantum subproblems. Specifically, BQA-MD combines block-QAOA-aware preprocessing in the QR domain, a standards-consistent blockwise 5G NR Gray-HUBO interface, an MMSE-induced dynamic regularized blockwise objective, and K-best candidate propagation. Within this framework, fixed-size block construction gives every local subproblem a uniform circuit width and parameter dimension, which in turn enables parameter-transfer QAOA as a practical realization strategy for structurally matched local subproblems.
Experiments are conducted on a $16\times16$ Rayleigh MIMO system with 16QAM using classical simulation of the quantum subroutine. The results show that the regularized blockwise detector improves upon its unregularized counterpart, validating the adopted blockwise objective and the block-QAOA-aware design rationale. They also show that the parameter-transfer QAOA detector nearly matches the regularized blockwise exhaustive reference and clearly outperforms direct-training QAOA in BER, thereby supporting parameter reuse as the preferred QAOA realization strategy within the proposed framework. In the tested setting, MMSE remains slightly better in the low-SNR region, whereas the parameter-transfer QAOA detector becomes highly competitive from the medium-SNR regime onward.
\end{abstract}


\section{Introduction}
\IEEEPARstart{M}{ultiple}-input multiple-output (MIMO) detection for large antenna arrays remains a fundamental yet challenging problem in modern wireless communications. Although maximum-likelihood (ML) detection achieves optimum error-rate performance, its search complexity grows rapidly with the antenna dimension and modulation order, making exact detection increasingly impractical in large-scale systems using QAM signaling. Classical near-ML detectors therefore commonly rely on QR-domain reformulation together with ordered successive interference cancellation and sorted QR decomposition (SQRD), or on breadth-first tree-search strategies such as K-best detection, to balance performance and complexity. More recently, group-sorted QR decomposition (GSQRD) has further shown that, in larger-scale MIMO systems, even the sorting stage itself becomes an important design bottleneck \cite{ref1,ref2,ref3}.

In parallel, quantum and quantum-inspired approaches to MIMO ML detection have developed rapidly in recent years. Cui \textit{et al.} introduced QAOA-based ML detection for binary-symbol MIMO systems, demonstrating that the detection problem can be encoded into a variational quantum circuit and solved in a hybrid quantum-classical manner \cite{ref4}. Norimoto \textit{et al.} then moved beyond purely quadratic formulations by proposing a quantum algorithm for higher-order unconstrained binary optimization (HUBO) and applying it to MIMO ML detection \cite{ref5}. Since 2024, the literature has further expanded toward large-scale QAOA-based MIMO detection, including several conference and journal studies for very large antenna settings, infinite-size analysis via the Sherrington-Kirkpatrick model with local field, recursive QAOA with majority voting and cost-restricted sampling, low-depth implementations aimed at reducing circuit depth and classical tuning overhead, and related QAOA-based MLD studies for MIMO-NOMA scenarios \cite{ref6,ref7,ref8,ref10,ref14}.

However, for large-scale MIMO systems with standardized QAM signaling, two additional issues must be handled carefully. The first is the modulation interface itself. Recent physics-inspired work on large-scale MIMO detection has emphasized that Gray-consistent modulation structure should be preserved rather than replaced by linearized low-order surrogates \cite{ref9}. The second issue is regularization. Singh \textit{et al.} showed that regularized Ising formulations can substantially improve near-optimal MIMO detection quality \cite{ref15}. In practical 5G NR systems, these considerations are especially relevant because the modulation mapper is standardized at the bit-to-symbol level in 3GPP TS 38.211, with explicit Gray-coded mappings for 16QAM, 64QAM, and higher-order constellations \cite{ref11}. Therefore, the optimization model should ideally remain consistent with the standardized bit-to-symbol mapping rather than replacing it with an artificial low-order surrogate for the sole convenience of obtaining a quadratic Ising or QUBO form. At the same time, the regularization result of \cite{ref15} is based on quantum-inspired Ising-machine dynamics rather than QAOA, and it does not address the combination of fixed-size blockwise decomposition, standards-consistent QAM modeling, and parameter transfer under limited local qubit budget.

Applying QAOA directly to a full high-dimensional MIMO detection instance still faces two practical bottlenecks. The first is the limited available qubit budget under current NISQ-era quantum resources, which constrains the size of the optimization instance that can be handled at once. The second is the need to fit QAOA parameters for a large number of local instances if no reusable structure is introduced. These two issues are closely coupled: unless the original detection problem is reorganized into local subproblems with controlled size and consistent structure, practical parameter reuse remains difficult to realize. Recent studies on QAOA parameter transfer suggest that optimized parameters can often be reused effectively across structurally related combinatorial instances \cite{ref12,ref13}. This observation motivates not only parameter reuse itself, but more fundamentally the design of a block-QAOA-aware detection chain in which local subproblems have standardized scale, interface, and parameter dimension.

Motivated by the above, this paper proposes a Block-QAOA-Aware MIMO Detector (BQA-MD) for large-scale MIMO. The core idea is to redesign the detection chain so that the resulting local optimization tasks become compatible with limited-qubit quantum processing in the current NISQ regime. Starting from the viewpoint of classical ordered SIC / SQRD, we generalize the conventional single-layer sorting proxy to a blockwise form and then specialize that blockwise rule to a fixed block size determined by the available qubit budget. This design choice is fundamentally different from GSQRD: whereas GSQRD is introduced to reduce sorting latency in larger-scale MIMO systems, the fixed block size in BQA-MD is chosen to standardize the dimension of local quantum subproblems and thereby create the structural precondition for downstream parameter reuse \cite{ref3,ref12,ref13}. On top of this foundation, we adopt a standards-consistent blockwise 5G NR Gray-HUBO interface, incorporate an MMSE-induced dynamic regularized blockwise objective, and integrate local QAOA outputs through K-best candidate propagation. Within this block-QAOA-aware framework, parameter transfer becomes the key realization strategy for reusing QAOA parameters across structurally matched local subproblems while preserving strong blockwise detection performance.

The main contributions of this paper are summarized as follows.
\begin{enumerate}
\item We propose a block-QAOA-aware preprocessing method inspired by classical ordered SIC / SQRD, where the conventional single-layer sorting proxy is generalized to a blockwise form and then specialized to a fixed block size according to the local qubit budget. This preprocessing is the structural basis of the proposed detector rather than an auxiliary implementation step.
\item We build a unified blockwise detection framework, namely BQA-MD, by combining fixed-size block decomposition, a standards-consistent 5G NR Gray-HUBO local modeling interface, an MMSE-induced dynamic regularized blockwise objective, and K-best candidate propagation for large-scale MIMO detection under limited local qubit budget.
\item On top of the block-QAOA-aware framework, we realize blockwise QAOA detection through parameter transfer and experimentally show in a $16\times16$ 16QAM setting that the parameter-transfer QAOA detector nearly matches the regularized blockwise exhaustive reference and clearly outperforms direct-training QAOA in BER.
\end{enumerate}

The remainder of this paper is organized as follows. Section II introduces the system model and formulates the large-scale MIMO detection problem under Gray-consistent QAM signaling. Section III presents the proposed BQA-MD, including the preprocessing stage, the blockwise 5G NR Gray-HUBO interface, the MMSE-induced dynamic regularized objective, the parameter-transfer QAOA solver, the K-best candidate propagation strategy, and the offline-online complexity discussion. Section IV reports the experimental results. Section V concludes the paper.

\section{System Model and Problem Formulation}

\subsection{System Model}

Consider an uplink or point-to-point narrowband MIMO system with $N_t$ transmit antennas and $N_r$ receive antennas. The received signal is modeled as
\begin{equation*}
\mathbf y=\mathbf H\mathbf x+\mathbf n,
\end{equation*}
where $\mathbf y\in\mathbb C^{N_r}$ is the received vector, $\mathbf H\in\mathbb C^{N_r\times N_t}$ is the channel matrix, $\mathbf x\in\mathcal X^{N_t}$ is the transmitted symbol vector, and $\mathbf n\sim\mathcal{CN}(\mathbf 0,\sigma^2\mathbf I_{N_r})$ is circularly symmetric complex Gaussian noise. The symbol alphabet $\mathcal X$ is an $M$-QAM constellation. For QAM modulation, each symbol carries
\begin{equation*}
m=\log_2 M
\end{equation*}
bits. Throughout this paper, the detector is formulated in the native complex-valued domain; binary variables are introduced only later for the blockwise Gray-HUBO representation of QAM symbols. In the proposed framework, the symbol-to-bit relation is later modeled according to the standardized Gray-coded mapping of 5G NR, while its explicit bit-level formulation is deferred to Section III-B.

Given $(\mathbf y,\mathbf H)$, the maximum-likelihood detector is
\begin{equation*}
\hat{\mathbf x}_{\rm ML}
=
\arg\min_{\mathbf x\in\mathcal X^{N_t}}
\|\mathbf y-\mathbf H\mathbf x\|_2^2.
\end{equation*}
Although optimal, this problem becomes increasingly difficult as $N_t$ and $M$ grow, which motivates a structured reformulation before introducing the proposed BQA-MD pipeline.

\subsection{QR-Domain Reformulation}

To obtain a detection model that is suitable for ordered or tree-based processing, we first consider a column-reordered channel matrix. Let $\mathbf \Pi\in\mathbb R^{N_t\times N_t}$ be a permutation matrix and define
\begin{equation*}
\mathbf H\mathbf \Pi=\mathbf Q\mathbf R,
\end{equation*}
where $\mathbf Q\in\mathbb C^{N_r\times N_t}$ has orthonormal columns and $\mathbf R\in\mathbb C^{N_t\times N_t}$ is upper triangular. Define the reordered transmit vector and the rotated received vector as
\begin{equation*}
\mathbf z=\mathbf \Pi^T\mathbf x,\qquad
\tilde{\mathbf y}=\mathbf Q^H\mathbf y.
\end{equation*}
Then the ML problem can be rewritten as
\begin{equation*}
\hat{\mathbf z}
=
\arg\min_{\mathbf z\in\mathcal X^{N_t}}
\|\tilde{\mathbf y}-\mathbf R\mathbf z\|_2^2,
\qquad
\hat{\mathbf x}=\mathbf \Pi \hat{\mathbf z}.
\end{equation*}
Since $\mathbf R$ is upper triangular, the reordered problem admits a sequential or tree-search interpretation. Expanding the metric yields
\begin{equation*}
\|\tilde{\mathbf y}-\mathbf R\mathbf z\|_2^2
=
\sum_{i=1}^{N_t}
\left|
\tilde y_i-\sum_{j=i}^{N_t}r_{i,j}z_j
\right|^2.
\end{equation*}
This form is the starting point for both classical ordered detection and the proposed block-structured quantum-aware detector.

\subsection{Symbolwise Sequential Interpretation}

The QR-domain upper-triangular form above admits the standard symbolwise sequential interpretation used in classical SIC and tree-search detectors. For a given suffix
\begin{equation*}
\mathbf z_{i+1:N_t}
=
\begin{bmatrix}
z_{i+1} & z_{i+2} & \cdots & z_{N_t}
\end{bmatrix}^T,
\end{equation*}
define the interference-reduced scalar observation
\begin{equation*}
\bar y_i
=
\tilde y_i-\sum_{j=i+1}^{N_t}r_{i,j}z_j .
\end{equation*}
Then the global metric can be written recursively as
\begin{equation*}
\|\tilde{\mathbf y}-\mathbf R\mathbf z\|_2^2
=
\sum_{i=1}^{N_t}
\left|
\bar y_i-r_{i,i}z_i
\right|^2.
\end{equation*}
Accordingly, conditioned on the suffix $\mathbf z_{i+1:N_t}$, the local symbolwise subproblem becomes
\begin{equation*}
\hat z_i
=
\arg\min_{z_i\in\mathcal X}
\left|
\bar y_i-r_{i,i}z_i
\right|^2.
\end{equation*}

For later use, define the cumulative suffix metric
\begin{equation*}
D_i(\mathbf z_{i:N_t})
=
\sum_{k=i}^{N_t}
\left|
\tilde y_k-\sum_{j=k}^{N_t}r_{k,j}z_j
\right|^2,
\end{equation*}
where
\begin{equation*}
\mathbf z_{i:N_t}
=
\begin{bmatrix}
z_i & z_{i+1} & \cdots & z_{N_t}
\end{bmatrix}^T.
\end{equation*}
Then
\begin{equation*}
D_i(\mathbf z_{i:N_t})
=
\left|
\bar y_i-r_{i,i}z_i
\right|^2
+
D_{i+1}(\mathbf z_{i+1:N_t}),
\end{equation*}
for $i=1,\ldots,N_t-1$, with terminal condition
\begin{equation*}
D_{N_t}(z_{N_t})
=
\left|
\tilde y_{N_t}-r_{N_t,N_t}z_{N_t}
\right|^2.
\end{equation*}
This is the conventional symbolwise QR-domain recursion underlying ordered SIC and K-best detection.

\subsection{Problem Formulation for BQA-MD}

The discussion above shows that the standard QR-domain formulation naturally leads to a symbolwise upper-triangular detection structure. This form is well aligned with classical SIC and tree-search detectors, but it is not yet tailored to the setting considered in this paper, where local QAOA processing is constrained by a limited qubit budget. The goal of this paper is therefore to preserve the QR-domain near-ML structure while reformulating the symbolwise recursion into a detector architecture that is compatible with fixed-size local quantum subproblems.

To this end, we seek a reformulated detection chain satisfying the following requirements:
\begin{enumerate}
\item the standard symbolwise upper-triangular detection structure should be reorganized into a fixed-size blockwise form compatible with QR-domain sequential detection;
\item the size of each local block should match the available local qubit budget;
\item each local block problem should admit a bit-level formulation consistent with the standardized Gray-coded modulation structure;
\item the resulting local quantum outputs should be combinable into a global near-ML search mechanism rather than being used in a purely greedy manner.
\end{enumerate}

Accordingly, the detection problem addressed in this paper can be stated as follows:

Given the received vector $\mathbf y$, the channel matrix $\mathbf H$, and an $M$-QAM alphabet $\mathcal X$,
find a QR-domain reformulation of the ML objective
\begin{equation*}
\min_{\mathbf x\in\mathcal X^{N_t}}\|\mathbf y-\mathbf H\mathbf x\|_2^2
\end{equation*}
such that the conventional symbolwise upper-triangular detection process is converted into a fixed-size blockwise upper-staircase structure whose local subproblems can be solved by a parameter-transfer QAOA strategy and whose candidate solutions can be integrated by a K-best propagation mechanism.

The construction of this blockwise upper-staircase detector is part of the proposed method and will be developed in Section III. Specifically, the next section introduces the block-QAOA-aware preprocessing rule, the blockwise 5G NR Gray-HUBO interface, the MMSE-induced dynamic regularized objective, the parameter-transfer QAOA solver, and the K-best candidate propagation mechanism.

\section{Proposed Block-QAOA-Aware MIMO Detector}
\subsection{Block-QAOA-aware preprocessing}

The preprocessing stage in BQA-MD reorganizes the original large-scale MIMO detection instance into a sequence of local subproblems that are explicitly compatible with downstream QAOA-based inference. The term ``block-QAOA-aware'' indicates that the block structure is chosen not only for QR-domain sequential detection, but also for limited-qubit local quantum processing and the parameter-transfer mechanism introduced later in Section III-D.

Following this design principle, we do not treat block construction as an independent classical grouping step appended after QR preprocessing. Instead, we start from the classical ordered SIC / SQRD viewpoint, generalize its single-layer sorting proxy to a blockwise form, and then specialize that blockwise rule to a fixed block size determined by the available local qubit budget. In this way, the preprocessing stage becomes the structural foundation of the entire BQA-MD framework rather than an auxiliary implementation detail.

Let
\begin{equation*}
\mathbf H=
\begin{bmatrix}
\mathbf h_1 & \mathbf h_2 & \cdots & \mathbf h_{N_t}
\end{bmatrix}
\in\mathbb C^{N_r\times N_t}
\end{equation*}
be the channel matrix. The preprocessing stage seeks a permutation matrix $\mathbf \Pi$ such that the reordered matrix $\mathbf H\mathbf \Pi$ admits a block structure that is favorable for the subsequent BQA-MD detector. To explain the proposed rule, we first recall the single-layer sorting mechanism underlying classical ordered SIC / SQRD.

At the $t$-th QR sorting step, let $\mathcal U_t\subseteq\{1,2,\ldots,N_t\}$ denote the set of unselected column indices, and let
\begin{equation*}
\mathbf Q_{t-1}=
\begin{bmatrix}
\mathbf q_1 & \mathbf q_2 & \cdots & \mathbf q_{t-1}
\end{bmatrix}
\end{equation*}
collect the orthonormal basis vectors already produced in the previous $t-1$ steps. Define the orthogonal projector onto the current selected subspace by
\begin{equation*}
\mathbf P_{t-1}=\mathbf Q_{t-1}\mathbf Q_{t-1}^H .
\end{equation*}
For any remaining column $\mathbf h_j$, $j\in\mathcal U_t$, its residual vector in the orthogonal complement of the selected subspace is
\begin{equation*}
\mathbf e_j^{(t)}=(\mathbf I-\mathbf P_{t-1})\mathbf h_j .
\end{equation*}
In classical ordered SIC / SQRD, a natural QR-internal sorting proxy is to rank the remaining columns according to their residual norms and select
\begin{equation}
\label{eq:sqrd_proxy}
j_t^\star
=
\arg\min_{j\in\mathcal U_t}
\left\|
(\mathbf I-\mathbf Q_{t-1}\mathbf Q_{t-1}^H)\mathbf h_j
\right\|_2^2 .
\end{equation}
This form is consistent with the classical SQRD viewpoint that ordering is embedded into the QR process itself rather than applied afterward; SQRD was introduced exactly to realize ordered SIC through sorted QR decomposition, while GSQRD later extended the idea to alleviate sorting latency in larger-scale MIMO systems \cite{ref1,ref3}. The design objective here is different: besides producing a favorable detection order, we must construct local subproblems whose scale can be matched to the subsequent quantum solver. To make the QR-internal sorting logic compatible with blockwise quantum processing, we therefore generalize the single-column proxy in \eqref{eq:sqrd_proxy} to a blockwise version. For any candidate index subset $B\subseteq \mathcal U_t$, define the corresponding column submatrix
\begin{equation*}
\mathbf H_B=
\begin{bmatrix}
\mathbf h_j
\end{bmatrix}_{j\in B}.
\end{equation*}
Its residual block matrix is
\begin{equation*}
\mathbf E_B^{(t)}
=
(\mathbf I-\mathbf P_{t-1})\mathbf H_B
=
(\mathbf I-\mathbf Q_{t-1}\mathbf Q_{t-1}^H)\mathbf H_B .
\end{equation*}
A natural scalarization of this residual block is its total residual energy
\begin{equation*}
\begin{aligned}
\left\|\mathbf E_B^{(t)}\right\|_F^2
&=
\left\|
(\mathbf I-\mathbf Q_{t-1}\mathbf Q_{t-1}^H)\mathbf H_B
\right\|_F^2 \\
&=
\operatorname{tr}\!\left(
\mathbf H_B^H(\mathbf I-\mathbf Q_{t-1}\mathbf Q_{t-1}^H)\mathbf H_B
\right).
\end{aligned}
\end{equation*}
Accordingly, the variable-size block proxy is defined as
\begin{equation}
\label{eq:block_proxy_var}
B_t^\star
=
\arg\min_{\substack{B\subseteq\mathcal U_t\\ 1\le |B|\le b_{\max}}}
\left\|
(\mathbf I-\mathbf Q_{t-1}\mathbf Q_{t-1}^H)\mathbf H_B
\right\|_F^2 .
\end{equation}
Equation~\eqref{eq:block_proxy_var} is the direct blockwise generalization of the single-layer SQRD proxy in \eqref{eq:sqrd_proxy}: when $|B|=1$, it reduces exactly to the classical residual-norm criterion. In this sense, the proposed block rule is not introduced heuristically from scratch, but is obtained by extending the ordered SIC / SQRD sorting logic from one column to one block. At this stage, however, the block size is still variable, so the resulting local subproblems are not yet suitable for standardized QAOA processing.

Equation~\eqref{eq:block_proxy_var} is only an intermediate form rather than the final one used in BQA-MD. The reason is twofold. First, the objective in \eqref{eq:block_proxy_var} is an accumulated residual energy, so allowing $|B|$ to vary tends to favor smaller candidate blocks. Second, and more importantly, the downstream QAOA solver requires local subproblems with a standardized dimension in order to maintain a uniform circuit width, a uniform parameter dimension, and a consistent local modeling interface. This is the key point that distinguishes the present framework from GSQRD: GSQRD is designed to reduce sorting latency, whereas here the block size is fixed in order to standardize the quantum subproblem scale and to create the precondition for parameter transfer \cite{ref3}.

Let $m=\log_2 M$ denote the number of bits carried by each $M$-QAM symbol, and let $n_{\rm avail}$ denote the available number of qubits for one local QAOA subproblem. Then the block size is fixed as
\begin{equation*}
b=
\left\lfloor
\frac{n_{\rm avail}}{m}
\right\rfloor .
\end{equation*}
In other words, each full-size block contains at most $b$ symbols so that its bit-level representation fits into the available qubit budget. With this choice, the final block-QAOA-aware preprocessing rule becomes
\begin{equation}
\label{eq:block_proxy_fixed}
B_t^\star
=
\arg\min_{\substack{B\subseteq\mathcal U_t\\ |B|=b}}
\left\|
(\mathbf I-\mathbf Q_{t-1}\mathbf Q_{t-1}^H)\mathbf H_B
\right\|_F^2 .
\end{equation}
Equation~\eqref{eq:block_proxy_fixed} is the actual preprocessing rule adopted in BQA-MD. It preserves the ordered SIC / SQRD-inspired residual-energy rationale while explicitly adapting the grouping scale to the quantum resource constraint. Consequently, each selected full-size block induces a local detection subproblem of the same nominal size, which is beneficial for circuit construction, local model standardization, and parameter reuse in the following QAOA stage. The transferability literature on QAOA supports this design choice by showing that optimized parameters can be effectively reused across structurally related instances \cite{ref12}.

After selecting $B_t^\star$, the corresponding columns are appended to the current permutation, removed from $\mathcal U_t$, and the QR update proceeds to the next stage. Repeating this process until all columns are assigned yields a block-reordered channel matrix
\begin{equation*}
\mathbf H\mathbf \Pi=
\begin{bmatrix}
\mathbf H_{B_1^\star} & \mathbf H_{B_2^\star} & \cdots & \mathbf H_{B_L^\star}
\end{bmatrix},
\end{equation*}
which is then factorized in QR form and passed to the subsequent blockwise detector. If $N_t$ is not an integer multiple of $b$, the final block can be handled as a smaller residual block without changing the fact that the dominant part of the detection chain is built from fixed-size local subproblems.

At this point, the detection structure departs from the conventional symbolwise QR-domain recursion of Section II. Let the reordered symbol vector be partitioned as
\begin{equation*}
\mathbf z=
\begin{bmatrix}
\mathbf z_1^T & \mathbf z_2^T & \cdots & \mathbf z_L^T
\end{bmatrix}^T,
\qquad
\mathbf z_\ell\in\mathcal X^{b_\ell},
\end{equation*}
where $b_\ell=b$ for all full-size blocks and $\sum_{\ell=1}^{L}b_\ell=N_t$. Correspondingly, define
\begin{equation*}
\tilde{\mathbf y}=
\begin{bmatrix}
\tilde{\mathbf y}_1^T & \tilde{\mathbf y}_2^T & \cdots & \tilde{\mathbf y}_L^T
\end{bmatrix}^T,
\end{equation*}
and partition the upper-triangular matrix $\mathbf R$ conformably as
\begin{equation*}
\mathbf R=
\begin{bmatrix}
\mathbf R_{1,1} & \mathbf R_{1,2} & \cdots & \mathbf R_{1,L}\\
\mathbf 0       & \mathbf R_{2,2} & \cdots & \mathbf R_{2,L}\\
\vdots          & \ddots          & \ddots & \vdots\\
\mathbf 0       & \cdots          & \mathbf 0 & \mathbf R_{L,L}
\end{bmatrix},
\end{equation*}
where each diagonal block $\mathbf R_{\ell,\ell}\in\mathbb C^{b_\ell\times b_\ell}$ is upper triangular. This is the block upper-staircase detection structure induced by the proposed preprocessing: the original symbolwise upper-triangular system is reorganized into a sequence of coupled blockwise subproblems.

Hence, after block-QAOA-aware preprocessing, the detector no longer operates symbol by symbol; instead, it proceeds block by block on the upper-staircase QR structure defined above. The corresponding conditional block metric, together with its Gray-HUBO realization, is introduced in the next subsection.

The purpose of this preprocessing stage is to make the reordered detection chain compatible with the later quantum solver. Accordingly, the adjective ``QAOA-aware'' in BQA-MD does not refer to a modified QAOA ansatz itself; it refers to a preprocessing rule whose block structure is explicitly chosen to match the size, interface, and transferability requirements of the downstream blockwise QAOA subproblems. Thus, block-QAOA-aware preprocessing forms the foundation of the proposed detector by converting a full large-scale MIMO instance into a structured family of local subproblems on which parameter-transfer QAOA can be applied consistently.

\subsection{Blockwise 5G NR Gray-HUBO interface}

Once the large-scale MIMO instance has been reorganized into fixed-size local blocks, the next step is to specify a local optimization interface that is consistent with the practical modulation rule and compatible with downstream QAOA processing. In BQA-MD, we adopt the standardized 5G NR Gray-coded bit-to-symbol mapping for that purpose, following prior HUBO-based MIMO formulations and recent physics-inspired work on large-scale MIMO modeling \cite{ref5,ref9,ref11}. The same binary interface is then used for both the unregularized and regularized block metrics.

For notational simplicity, we first consider a full-size block of length $b$; the final residual block, if any, can be handled analogously. Under the block upper-staircase structure established in Section III-A, and for a given suffix $\mathbf z_{\ell+1},\ldots,\mathbf z_L$, define the blockwise interference-reduced observation as
\begin{equation}
\label{eq:block_obs}
\bar{\mathbf y}_\ell
=
\tilde{\mathbf y}_\ell
-
\sum_{j=\ell+1}^{L}\mathbf R_{\ell,j}\mathbf z_j.
\end{equation}
Then the $\ell$-th conditional block metric is
\begin{equation}
\label{eq:block_metric}
\hat{\mathbf z}_\ell
=
\arg\min_{\mathbf z_\ell\in\mathcal X^{b}}
\left\|
\bar{\mathbf y}_\ell-\mathbf R_{\ell,\ell}\mathbf z_\ell
\right\|_2^2 ,
\end{equation}
where $\mathbf z_\ell=[z_{\ell,1},z_{\ell,2},\ldots,z_{\ell,b}]^T\in\mathcal X^b$, $\bar{\mathbf y}_\ell\in\mathbb C^b$, and $\mathbf R_{\ell,\ell}\in\mathbb C^{b\times b}$.

Let
\begin{equation*}
m=\log_2 M
\end{equation*}
denote the number of bits per $M$-QAM symbol. For the $t$-th symbol in block $\ell$, define the corresponding bit vector as
\begin{equation*}
\begin{aligned}
\mathbf b_{\ell,t}
&=
\begin{bmatrix}
b_{\ell,t,1} & b_{\ell,t,2} & \cdots & b_{\ell,t,m}
\end{bmatrix}^T,\\
\mathbf b_{\ell,t}
&\in\{0,1\}^{m},\qquad t=1,2,\ldots,b.
\end{aligned}
\end{equation*}
Then the symbol $z_{\ell,t}$ is determined by the 5G NR Gray-coded modulation mapping
\begin{equation*}
z_{\ell,t}=g_M(\mathbf b_{\ell,t}),
\end{equation*}
where $g_M:\{0,1\}^m\rightarrow\mathcal X$ denotes the standardized bit-to-symbol mapper specified in 3GPP TS 38.211.

To connect the detection problem with higher-order binary optimization, we introduce Ising spin variables
\begin{equation}
\label{eq:spin_def}
s_{\ell,t,r}=1-2b_{\ell,t,r}\in\{-1,+1\},
\qquad r=1,2,\ldots,m.
\end{equation}
Let
\begin{equation*}
\mathbf s_{\ell,t}
=
\begin{bmatrix}
s_{\ell,t,1} & s_{\ell,t,2} & \cdots & s_{\ell,t,m}
\end{bmatrix}^T
\in\{-1,+1\}^{m}.
\end{equation*}
Then the Gray-coded mapper can equivalently be written as
\begin{equation*}
z_{\ell,t}=\tilde g_M(\mathbf s_{\ell,t}),
\end{equation*}
where $\tilde g_M(\cdot)$ is the same 5G NR modulation rule expressed in spin variables.

Next, collect all spins in block $\ell$ as
\begin{equation*}
\mathbf s_\ell
=
\begin{bmatrix}
\mathbf s_{\ell,1}^T & \mathbf s_{\ell,2}^T & \cdots & \mathbf s_{\ell,b}^T
\end{bmatrix}^T
\in\{-1,+1\}^{bm},
\end{equation*}
and define the blockwise Gray mapping
\begin{equation}
\label{eq:block_gray_map}
\mathbf z_\ell=\mathbf g_M(\mathbf s_\ell)
=
\begin{bmatrix}
\tilde g_M(\mathbf s_{\ell,1}) &
\tilde g_M(\mathbf s_{\ell,2}) &
\cdots &
\tilde g_M(\mathbf s_{\ell,b})
\end{bmatrix}^T .
\end{equation}
Substituting \eqref{eq:spin_def} into the local detection metric \eqref{eq:block_metric} yields
\begin{equation}
\label{eq:block_cost_qr}
f_\ell(\mathbf s_\ell)
=
\left\|
\bar{\mathbf y}_\ell-\mathbf R_{\ell,\ell}\mathbf g_M(\mathbf s_\ell)
\right\|_2^2 .
\end{equation}

For later use, define
\begin{equation*}
\mathbf G_\ell=\mathbf R_{\ell,\ell}^H\mathbf R_{\ell,\ell},
\qquad
\mathbf u_\ell=\mathbf R_{\ell,\ell}^H\bar{\mathbf y}_\ell .
\end{equation*}
Then \eqref{eq:block_cost_qr} can be rewritten as
\begin{equation}
\label{eq:block_cost_qr_quad}
f_\ell(\mathbf s_\ell)
=
\mathbf g_M(\mathbf s_\ell)^H\mathbf G_\ell \mathbf g_M(\mathbf s_\ell)
-2\operatorname{Re}\!\left\{\mathbf u_\ell^H\mathbf g_M(\mathbf s_\ell)\right\}
+\bar{\mathbf y}_\ell^H\bar{\mathbf y}_\ell .
\end{equation}

Equation~\eqref{eq:block_cost_qr_quad} gives the QR-domain block cost before regularization. Since the Gray-HUBO mapping itself is independent of whether regularization is added, the same variable definition can be retained when the local objective is further augmented. The corresponding MMSE-induced dynamic regularization is introduced in the next subsection.

\subsection{MMSE-induced dynamic regularized blockwise objective}

In the present detector, the regularized local objective is formed by anchoring the blockwise symbol hypothesis to the reordered hard MMSE reference on the same block. Let $\mathbf z_{\ell}^{\rm MMSE}\in\mathcal X^b$ denote that blockwise MMSE reference. The regularized cost is written as
\begin{equation}
\label{eq:block_cost_reg}
f_\ell^{\rm reg}(\mathbf s_\ell;\rho)
=
f_\ell(\mathbf s_\ell)
+
\lambda(\rho)
\left\|
\mathbf g_M(\mathbf s_\ell)-\mathbf z_{\ell}^{\rm MMSE}
\right\|_2^2,
\end{equation}
where $\rho$ denotes the nominal SNR index of the current sample. The regularization weight is chosen by the SNR-dependent schedule
\begin{equation}
\label{eq:lambda_schedule}
\lambda(\rho)
=
\lambda_{\min}
+
\frac{\lambda_{\max}-\lambda_{\min}}
{1+\exp\!\big(\kappa(\rho-\rho_0)\big)}.
\end{equation}
Thus, stronger regularization is applied in the low-SNR region and weaker regularization is used in the high-SNR region. In the experiments of Section IV, $(\lambda_{\min},\lambda_{\max},\rho_0,\kappa)$ are fixed globally and shared by all blockwise realizations using regularization.

At this point, the key observation is that each entry of $\mathbf g_M(\mathbf s_\ell)$ is a multilinear polynomial of the spin variables under Gray-coded QAM modulation. Therefore, both the unregularized objective in \eqref{eq:block_cost_qr_quad} and the regularized objective in \eqref{eq:block_cost_reg} admit higher-order polynomial representations over $\mathbf s_\ell$. To keep the notation compact, let $\tilde f_\ell(\mathbf s_\ell;\rho)$ denote the local objective actually used in the detector, i.e., $\tilde f_\ell=f_\ell$ in the unregularized case and $\tilde f_\ell=f_\ell^{\rm reg}$ in the regularized case. Then
\begin{equation*}
\begin{aligned}
\tilde f_\ell(\mathbf s_\ell;\rho)
&=
c_{\ell,\emptyset}
+
\sum_{\substack{\emptyset\neq S\\S\subseteq\{1,\ldots,bm\}}}
c_{\ell,S}\prod_{k\in S}s_{\ell,k}.
\end{aligned}
\end{equation*}
where $\mathbf s_\ell\in\{-1,+1\}^{bm}$, $c_{\ell,\emptyset}\in\mathbb R$ is a constant term, and $c_{\ell,S}\in\mathbb R$ are real-valued coefficients determined by $\mathbf R_{\ell,\ell}$, $\bar{\mathbf y}_\ell$, the adopted Gray mapping, and, when regularization is enabled, $\lambda(\rho)$ together with $\mathbf z_{\ell}^{\rm MMSE}$. Since the constant term does not affect the optimizer, the $\ell$-th local block detection problem is finally written as
\begin{equation}
\label{eq:gray_hubo_subproblem}
\hat{\mathbf s}_\ell
=
\arg\min_{\mathbf s_\ell\in\{-1,+1\}^{bm}}
\sum_{\emptyset\neq S\subseteq\{1,2,\ldots,bm\}}
c_{\ell,S}\prod_{k\in S}s_{\ell,k}.
\end{equation}
Equation~\eqref{eq:gray_hubo_subproblem} defines the fixed-size Gray-HUBO subproblem associated with each block in BQA-MD. Its coefficients are determined by the QR-domain block quantities, the adopted Gray mapping, and, when regularization is enabled, the dynamic weight $\lambda(\rho)$ together with the MMSE reference. The modulation-specific explicit forms follow directly from the 5G NR specification and related prior derivations \cite{ref5,ref9,ref11}. In the next subsection, this fixed-size binary subproblem is solved by the parameter-transfer QAOA realization of BQA-MD.

\subsection{Parameter-transfer QAOA solver}

Given the fixed-size blockwise local QAOA subproblem induced by the adopted metric and Gray-HUBO interface, the next step of BQA-MD is to solve each local binary optimization problem by QAOA. Because Section III-A standardizes the dominant local blocks to the same size and parameter dimension, parameter transfer becomes the primary realization of this local QAOA stage, while direct training is retained only as a comparison mode in the experiments. Let
\begin{equation*}
q=bm
\end{equation*}
denote the number of binary variables, and hence the number of qubits, associated with a full-size block. For the $\ell$-th block, each computational basis state $|\mathbf s_\ell\rangle$ is identified with a spin assignment
\begin{equation*}
\mathbf s_\ell\in\{-1,+1\}^{q}.
\end{equation*}
Accordingly, the local objective in \eqref{eq:gray_hubo_subproblem} can be encoded into a diagonal cost Hamiltonian
\begin{equation*}
\hat H_{C,\ell}
=
\sum_{\emptyset\neq S\subseteq\{1,2,\ldots,q\}}
c_{\ell,S}\prod_{k\in S}\hat Z_k ,
\end{equation*}
where $\hat Z_k$ is the Pauli-$Z$ operator acting on the $k$-th qubit. Since the constant term $c_{\ell,\emptyset}$ does not affect the optimizer, it is omitted from $\hat H_{C,\ell}$. By construction, for any computational basis state $|\mathbf s_\ell\rangle$, the eigenvalue of $\hat H_{C,\ell}$ equals the nonconstant part of the blockwise cost in \eqref{eq:gray_hubo_subproblem}.

In this work, we adopt the standard unconstrained QAOA form with a transverse-field mixer
\begin{equation*}
\hat H_M
=
\sum_{k=1}^{q}\hat X_k ,
\end{equation*}
where $\hat X_k$ is the Pauli-$X$ operator on qubit $k$. Starting from the uniform superposition state
\begin{equation*}
|+\rangle^{\otimes q},
\end{equation*}
the $p$-layer QAOA ansatz for the $\ell$-th block is defined as
\begin{equation*}
|\psi_\ell(\boldsymbol\gamma_\ell,\boldsymbol\beta_\ell)\rangle
=
\left(
\prod_{r=1}^{p}
e^{-i\beta_{\ell,r}\hat H_M}
e^{-i\gamma_{\ell,r}\hat H_{C,\ell}}
\right)
|+\rangle^{\otimes q},
\end{equation*}
where
\begin{equation*}
\boldsymbol\gamma_\ell=
[\gamma_{\ell,1},\gamma_{\ell,2},\ldots,\gamma_{\ell,p}]^T,\qquad
\boldsymbol\beta_\ell=
[\beta_{\ell,1},\beta_{\ell,2},\ldots,\beta_{\ell,p}]^T
\end{equation*}
are the variational parameters. Let
\begin{equation*}
\boldsymbol\theta_\ell=
\begin{bmatrix}
\boldsymbol\gamma_\ell^T & \boldsymbol\beta_\ell^T
\end{bmatrix}^T
\in\mathbb R^{2p}
\end{equation*}
collect all circuit parameters. The corresponding classical objective is
\begin{equation*}
F_\ell(\boldsymbol\theta_\ell)
=
\langle\psi_\ell(\boldsymbol\theta_\ell)|
\hat H_{C,\ell}
|\psi_\ell(\boldsymbol\theta_\ell)\rangle .
\end{equation*}
The locally optimized QAOA parameters are then given by
\begin{equation}
\label{eq:qaoa_param_opt}
\boldsymbol\theta_\ell^\star
=
\arg\min_{\boldsymbol\theta_\ell\in\mathbb R^{2p}}
F_\ell(\boldsymbol\theta_\ell).
\end{equation}

A direct implementation of \eqref{eq:qaoa_param_opt} yields a per-block direct-training realization. In BQA-MD, however, the preferred realization is parameter transfer, motivated by recent QAOA transferability results showing that optimized parameters can often be reused effectively across structurally related combinatorial instances \cite{ref12,ref13}. In the present framework, this observation is not used as an auxiliary trick; it follows directly from the fixed-size block construction established in Section III-A.

BQA-MD is enabled by the fixed-size block construction of Section III-A. Because each full-size block contains exactly $b$ symbols, every local subproblem has the same number of binary variables $q=bm$, the same circuit width, and the same parameter dimension $2p$. Therefore, the same parameter format can be reused across structurally matched blocks without any dimension mismatch. In the present framework, this transfer is realized through an offline template bank indexed by SNR. For each SNR grid point $\rho$, let
\begin{equation*}
\Theta_\rho
=
\left\{
\boldsymbol\theta_{\rho,1}^{\rm ref},
\boldsymbol\theta_{\rho,2}^{\rm ref},
\ldots,
\boldsymbol\theta_{\rho,K_{\rm temp}}^{\rm ref}
\right\}
\end{equation*}
denote the stored reference parameter set, where $K_{\rm temp}$ is the number of transferred templates. In the experiments of this paper, we use $K_{\rm temp}=4$.

For a local block observed at SNR $\rho$, each transferred template provides a candidate parameter setting for the corresponding QAOA circuit. In the parameter-transfer realization considered in this paper, the transferred template is used directly, namely
\begin{equation*}
\boldsymbol\theta_{\ell,k}^\star
=
\boldsymbol\theta_{\rho,k}^{\rm ref},
\qquad k=1,2,\ldots,K_{\rm temp},
\end{equation*}
that is, the local QAOA circuit is evaluated without any additional per-instance parameter update. For comparison, a direct-training realization is obtained by locally optimizing \eqref{eq:qaoa_param_opt} for each block without using transferred parameters.

After obtaining $\boldsymbol\theta_{\ell,k}^\star$, the corresponding circuit is measured $N_{\rm sh}$ times in the computational basis. Let
\begin{equation*}
\mathcal M_{\ell,k}
=
\left\{
\mathbf s_{\ell,k}^{(a)}
\right\}_{a=1}^{N_{\rm sh}}
\end{equation*}
denote the sampled bitstrings for the $k$-th template. From the resulting measurement outcomes, we extract distinct local candidates and their empirical sampling frequencies. For each candidate $\mathbf s_{\ell,k}^{(a)}$, the corresponding symbol vector is recovered by the blockwise Gray mapping
\begin{equation*}
\mathbf z_{\ell,k}^{(a)}=\mathbf g_M(\mathbf s_{\ell,k}^{(a)}),
\end{equation*}
and its exact adopted local blockwise metric is evaluated as $\Delta_{\ell,k}^{(a)}$.
For each transferred template, we obtain a local candidate list
\begin{equation*}
\mathcal C_{\ell,k}
=
\left\{
\big(
\mathbf s_{\ell,k}^{(a)},
\mathbf z_{\ell,k}^{(a)},
\Delta_{\ell,k}^{(a)},
\hat\pi_{\ell,k}^{(a)}
\big)
\right\}_{a=1}^{T}.
\end{equation*}

The present implementation further adopts block mixing across the transferred templates. That is, for the $\ell$-th block we aggregate the candidate sets produced by all templates and keep the best $T$ distinct candidates according to the exact local metric:
\begin{equation*}
\mathcal C_\ell
=
\operatorname{Top}\text{-}T
\left(
\bigcup_{k=1}^{K_{\rm temp}}
\mathcal C_{\ell,k}
\right).
\end{equation*}
This list is the final local output of the parameter-transfer QAOA solver and will be integrated into the global K-best propagation procedure in the next subsection. In this way, the block-QAOA-aware preprocessing, the fixed-size Gray-HUBO local model, and the parameter-transfer QAOA solver form a single detector pipeline, while the direct-training realization remains a useful reference mode for evaluating the effect of transfer.

\subsection{K-best candidate propagation}

The parameter-transfer QAOA solver described in Section III-D produces, for each conditioned local block, a finite list of candidate local solutions. These local outputs are integrated by a K-best propagation mechanism in the QR domain, consistent with the classical breadth-first interpretation of K-best MIMO detection in which only the $K$ nodes with the smallest accumulated partial Euclidean distances (PEDs) are retained at each search level \cite{ref2}.

Using the block partition introduced in Section III-A, the reordered symbol vector is written as
\begin{equation*}
\mathbf z=
\begin{bmatrix}
\mathbf z_1^T & \mathbf z_2^T & \cdots & \mathbf z_L^T
\end{bmatrix}^T,
\end{equation*}
and the corresponding cumulative path metric for a candidate suffix $\mathbf z_{\ell:L}:=(\mathbf z_\ell,\mathbf z_{\ell+1},\ldots,\mathbf z_L)$ is
\begin{equation*}
D_\ell(\mathbf z_{\ell:L})
=
\sum_{i=\ell}^{L}
\left\|
\bar{\mathbf y}_i-\mathbf R_{i,i}\mathbf z_i
\right\|_2^2 .
\end{equation*}
Equivalently, the recursion
\begin{equation*}
D_\ell(\mathbf z_{\ell:L})
=
\left\|
\bar{\mathbf y}_\ell-\mathbf R_{\ell,\ell}\mathbf z_\ell
\right\|_2^2
+
D_{\ell+1}(\mathbf z_{\ell+1:L})
\end{equation*}
holds for $\ell=1,\ldots,L-1$, with
\begin{equation*}
D_L(\mathbf z_L)
=
\left\|
\tilde{\mathbf y}_L-\mathbf R_{L,L}\mathbf z_L
\right\|_2^2 .
\end{equation*}
These accumulated distances play the same role as the PEDs in classical QR-tree detection and are used here as the global ranking metric for blockwise candidate propagation.

BQA-MD uses different metrics for local candidate generation and global path pruning. At the local level, the QAOA solver operates on the adopted blockwise objective described in Sections III-B and III-C, which may include regularization. At the global level, candidate paths are ranked by accumulated QR-domain PED, which provides the consistent path metric for breadth-first propagation across blocks.

We solve the blocks in backward order, i.e., from $\ell=L$ to $\ell=1$, so that each new block extends an already constructed suffix. Suppose that at level $\ell+1$, the candidate suffix list
\begin{equation*}
\mathcal L_{\ell+1}
=
\left\{
\big(
\mathbf z_{\ell+1:L}^{(k)},
D_{\ell+1}^{(k)}
\big)
\right\}_{k=1}^{|\mathcal L_{\ell+1}|}
\end{equation*}
has already been retained. For each parent suffix $k$, the corresponding interference-reduced observation is
\begin{equation*}
\bar{\mathbf y}_\ell^{(k)}
=
\tilde{\mathbf y}_\ell
-
\sum_{j=\ell+1}^{L}\mathbf R_{\ell,j}\mathbf z_j^{(k)} .
\end{equation*}
Using this conditioned local problem, the parameter-transfer QAOA solver generates the parent-dependent local candidate list
\begin{equation*}
\mathcal C_\ell^{(k)}
=
\left\{
\big(
\mathbf s_\ell^{(k,a)},
\mathbf z_\ell^{(k,a)},
\Delta_\ell^{(k,a)},
\hat\pi_\ell^{(k,a)}
\big)
\right\}_{a=1}^{T},
\end{equation*}
where $\Delta_\ell^{(k,a)}$ denotes the exact adopted local blockwise metric used inside the local solver, and $\hat\pi_\ell^{(k,a)}$ is the corresponding empirical sampling frequency. These local candidates are then propagated block by block through a breadth-first tree search.

At the last block, the initial list is formed directly from the local candidates:
\begin{equation*}
\mathcal L_L
=
\operatorname{TopK}_{D_L}
\left(
\left\{
\big(
\mathbf z_L^{(a)},
D_L^{(a)}
\big)
\right\}_{a=1}^{T}
\right).
\end{equation*}
\begin{equation*}
D_L^{(a)}
=
\left\|
\tilde{\mathbf y}_L-\mathbf R_{L,L}\mathbf z_L^{(a)}
\right\|_2^2 .
\end{equation*}
Here, $\operatorname{TopK}_{D_L}(\cdot)$ denotes selecting the $K$ candidates with the smallest accumulated metric.

For each parent candidate and each local candidate $a\in\{1,\ldots,T\}$ from $\mathcal C_\ell^{(k)}$, a new child candidate is constructed as
\begin{equation*}
\mathbf z_{\ell:L}^{(k,a)}
=
\begin{bmatrix}
(\mathbf z_\ell^{(k,a)})^T & (\mathbf z_{\ell+1:L}^{(k)})^T
\end{bmatrix}^T .
\end{equation*}
Its accumulated path metric is computed by
\begin{equation*}
D_\ell^{(k,a)}
=
D_{\ell+1}^{(k)}
+
\delta_\ell^{(k,a)},
\end{equation*}
where
\begin{equation*}
\delta_\ell^{(k,a)}
=
\left\|
\bar{\mathbf y}_\ell^{(k)}-\mathbf R_{\ell,\ell}\mathbf z_\ell^{(k,a)}
\right\|_2^2 ,
\end{equation*}
which is the QR-domain PED increment associated with parent $k$ and child $a$.
Thus, every parent candidate generates up to $T$ child candidates, and all children together form the expanded set
\begin{equation*}
\mathcal E_\ell
=
\bigcup_{k=1}^{|\mathcal L_{\ell+1}|}
\left\{
\big(
\mathbf z_{\ell:L}^{(k,a)},
D_\ell^{(k,a)}
\big)
\right\}_{a=1}^{T}.
\end{equation*}
The propagated list at level $\ell$ is then obtained by breadth-first pruning:
\begin{equation}
\label{eq:kbest_propagation}
\mathcal L_\ell
=
\operatorname{TopK}_{D_\ell}(\mathcal E_\ell).
\end{equation}

Equation~\eqref{eq:kbest_propagation} is the core propagation rule of BQA-MD. It keeps the $K$ most promising blockwise partial paths according to their accumulated QR-domain metrics. In particular, if one always retains only the best local candidate and only one global path, i.e.,
\begin{equation*}
T=1,\qquad K=1,
\end{equation*}
then BQA-MD degenerates into a greedy blockwise successive detector. By contrast, choosing
\begin{equation*}
T>1,\qquad K>1
\end{equation*}
allows the detector to preserve multiple near-best hypotheses, which is the same mechanism by which K-best search mitigates error propagation in classical tree-based MIMO detection \cite{ref2}.

The final hard-output decision of BQA-MD is obtained from the best complete path in $\mathcal L_1$:
\begin{equation*}
\hat{\mathbf z}
=
\arg\min_{\mathbf z_{1:L}\in\mathcal L_1} D_1(\mathbf z_{1:L}),
\qquad
\hat{\mathbf x}=\mathbf \Pi \hat{\mathbf z}.
\end{equation*}
Hence, BQA-MD combines local quantum optimization with global breadth-first candidate selection: QAOA produces parent-conditioned local block candidates, while K-best propagation enforces a global ranking in terms of accumulated QR-domain PEDs.

\subsection{Offline-Online Complexity Discussion}
\label{subsec:complexity_discussion}

The role of parameter transfer in BQA-MD is to change the computational structure of local QAOA processing rather than to claim a backend-independent wall-clock advantage under every simulator or hardware setting. Actual runtime depends strongly on the adopted optimizer, circuit compilation flow, shot interface, and whether the backend is a classical simulator or a physical quantum device. Therefore, the present paper summarizes the detector complexity at the structural level by separating shared preprocessing, online detection, and offline template construction.

Let
\begin{equation*}
L=\left\lceil \frac{N_t}{b} \right\rceil,
\qquad
q=b\log_2 M
\end{equation*}
denote the number of detection blocks and the number of qubits associated with a full-size local QAOA subproblem, respectively. Because the backward blockwise K-best search retains at most $K$ parent paths after each level, the number of conditioned local subproblems solved for one received sample is bounded by
\begin{equation*}
N_{\rm loc}\le 1+(L-1)K = O(LK).
\end{equation*}
The last block contributes one local subproblem, whereas each of the remaining $L-1$ levels contributes at most $K$ parent-conditioned local problems.

The QR-domain preprocessing and block construction are shared by all BQA-MD realizations. In particular, the associated QR factorization is polynomial in antenna dimension and contributes order $N_t^3$ in the square case $N_r=N_t$. The breadth-first blockwise propagation is also common to both direct-training and transfer-based realizations: each level expands at most $KT$ child paths and keeps the best $K$, which yields a list-management cost of order
\begin{equation*}
O\!\left(LKT\log(KT)\right),
\end{equation*}
while the corresponding QR-domain PED updates contribute order
\begin{equation*}
O\!\left(LKTb^2\right).
\end{equation*}
These shared classical terms do not distinguish parameter transfer from direct training.

To compare the local QAOA stages, let $\mathcal C_{\rm inf}(q,p,N_{\rm sh})$ denote the cost of one fixed-parameter $q$-qubit, $p$-layer QAOA inference call with $N_{\rm sh}$ shots on the adopted backend, and let $\mathcal C_{\rm opt}(q,p,N_{\rm sh},I_{\rm on})$ denote the cost of one online parameter optimization with $I_{\rm on}$ optimizer iterations for the same local problem. Then the direct-training realization incurs the online local cost
\begin{equation*}
\mathcal C_{\rm direct}^{\rm on}
=
O\!\left(
N_{\rm loc}\,\mathcal C_{\rm opt}(q,p,N_{\rm sh},I_{\rm on})
\right),
\end{equation*}
because every conditioned local block requires per-instance online parameter fitting. By contrast, single-template parameter transfer replaces this stage by
\begin{equation*}
\mathcal C_{\rm transfer}^{\rm on}
=
O\!\left(
N_{\rm loc}\,\mathcal C_{\rm inf}(q,p,N_{\rm sh})
\right),
\end{equation*}
which removes per-instance online variational optimization. In the top-$K_{\rm temp}$ block-mixing realization adopted in this paper, the online local cost becomes
\begin{equation*}
\mathcal C_{\rm transfer,mix}^{\rm on}
=
O\!\left(
N_{\rm loc}\,K_{\rm temp}\,\mathcal C_{\rm inf}(q,p,N_{\rm sh})
\right),
\end{equation*}
plus the linear overhead of aggregating and rescoring the candidates produced by the transferred templates. Hence, parameter transfer changes the online burden from optimization-dominated local processing to inference-dominated local processing, while multi-template block mixing deliberately trades additional online template evaluations for improved robustness.

The transfer realization also introduces an offline template-bank cost. Let $G$ denote the number of SNR grid points used to build the template bank and let $N_{\rm ref}$ denote the number of reference training instances per grid point. If $K_{\rm temp}$ templates are retained for each SNR and each template stores a $2p$-dimensional QAOA parameter vector, then the offline construction cost scales as
\begin{equation*}
O\!\left(
G\,N_{\rm ref}\,\mathcal C_{\rm opt}(q,p,N_{\rm sh},I_{\rm off})
\right),
\end{equation*}
where $I_{\rm off}$ is the offline optimization budget, and the template-bank storage scales as
\begin{equation*}
O(GK_{\rm temp}p).
\end{equation*}
This cost is paid once and can then be reused across many received samples.

Under this decomposition, the main complexity advantage claimed for parameter transfer is not a universal simulator-side runtime reduction, but the relocation of parameter-learning effort from online per-instance optimization to offline template construction. This is also why Section IV focuses on BER rather than raw wall-clock time: on a classical state-vector simulator, measured runtime can be dominated by backend-specific implementation details that do not faithfully represent the intended offline-online separation of the transfer strategy.

\subsection{Algorithm summary}

Algorithm~\ref{alg:bqa_md} summarizes the per-sample workflow of BQA-MD from block-QAOA-aware preprocessing to global path selection. It corresponds to the parameter-transfer realization of BQA-MD used in the experiments of Section IV.

\begin{algorithm}[t]
\caption{BQA-MD Detection}
\label{alg:bqa_md}
\begin{algorithmic}[1]
\REQUIRE $\mathbf y,\mathbf H$, SNR index $\rho$, $(b,p,T,K,N_{\rm sh})$, transferred template bank $\{\Theta_\rho\}$.
\ENSURE Detected symbol vector $\hat{\mathbf x}$.
\STATE Apply block-QAOA-aware preprocessing and QR decomposition to obtain $\mathbf \Pi,\mathbf R,\tilde{\mathbf y}$.
\STATE Construct the blockwise local Gray-HUBO model and initialize $\mathcal L_{L+1}\leftarrow\{(\emptyset,0)\}$.
\FOR{$\ell=L,L-1,\ldots,1$}
\FOR{each parent path retained in $\mathcal L_{\ell+1}$}
\STATE Build the conditioned local blockwise subproblem for block $\ell$.
\STATE Run QAOA inference using the transferred templates in $\Theta_\rho$.
\STATE Extract local candidates by block mixing and keep the best $T$ candidates.
\STATE Expand child paths and evaluate their QR-domain PED increments.
\ENDFOR
\STATE Retain the best $K$ expanded paths to form $\mathcal L_\ell$.
\ENDFOR
\STATE Select the best full path in $\mathcal L_1$ to obtain $\hat{\mathbf z}$ and return $\hat{\mathbf x}=\mathbf \Pi \hat{\mathbf z}$.
\end{algorithmic}
\end{algorithm}

\section{Experimental Results}

\subsection{Simulation Setup}
We evaluate uncoded detection performance on a narrowband flat-fading MIMO link with
\begin{equation*}
N_t=N_r=16,
\end{equation*}
under i.i.d. Rayleigh fading and additive white Gaussian noise. The modulation is 16QAM with 5G NR Gray-consistent symbol mapping. The tested SNR range is
\begin{equation*}
\{0,2,4,\ldots,30\}\ \mathrm{dB}.
\end{equation*}
For each SNR point, 100 independent channel uses are generated and the average bit error rate (BER) is reported. All random generation and detector runs use fixed seeds for reproducibility.

All experiments are performed by classical simulation. Specifically, local QAOA subproblems are executed on the state-vector simulator backend using the MindSpore Quantum framework~\cite{ref16}, rather than on quantum hardware. Therefore, simulator-side wall-clock time is not used here as a principal evaluation metric. Since the actual execution time of QAOA depends strongly on the adopted optimizer, stopping criterion, and hardware backend, Section IV focuses on BER behavior rather than raw runtime, while the corresponding offline-online complexity discussion is given in Section~\ref{subsec:complexity_discussion}. The source code and data required to reproduce the numerical results are publicly available at \url{https://gitee.com/beastsenior/bqa_md}.

The compared detectors are MF, MMSE, classical K-best, and four variants of the proposed blockwise detector: 1) the regularized blockwise exhaustive detector; 2) the unregularized blockwise exhaustive detector; 3) the regularized blockwise QAOA detector with direct training; and 4) the regularized blockwise QAOA detector with parameter transfer, top-4 templates, and block mixing. The common BQA-MD configuration is summarized in Table~\ref{tab:v6_setup}.

\begin{table}[t]
\caption{Default Experimental Configuration for the $16\times16$ 16QAM BQA-MD Study}
\label{tab:v6_setup}
\centering
\renewcommand{\arraystretch}{1.12}
\begin{tabular}{|c|c|}
\hline
Parameter & Value \\
\hline
$N_t,N_r$ & $16,16$ \\
\hline
Modulation & 16QAM \\
\hline
SNR grid & $0{:}2{:}30$ dB \\
\hline
Block size $b$ & 2 \\
\hline
Qubit budget $q=b\log_2 M$ & 8 \\
\hline
QAOA depth $p$ & 4 \\
\hline
Shots & 1024 \\
\hline
Local list width $T$ & 4 \\
\hline
Global K-best width $K$ & 4 \\
\hline
Regularization & dynamic $\lambda(\rho)$ \\
\hline
$(\lambda_{\min},\lambda_{\max},\rho_0,\kappa)$ & $(0.005,0.45,13,0.55)$ \\
\hline
Transfer bank size & 4 templates / SNR \\
\hline
\end{tabular}
\end{table}

\subsection{Overall BER Comparison}
\begin{figure}[!t]
\centering
\includegraphics[width=0.98\linewidth]{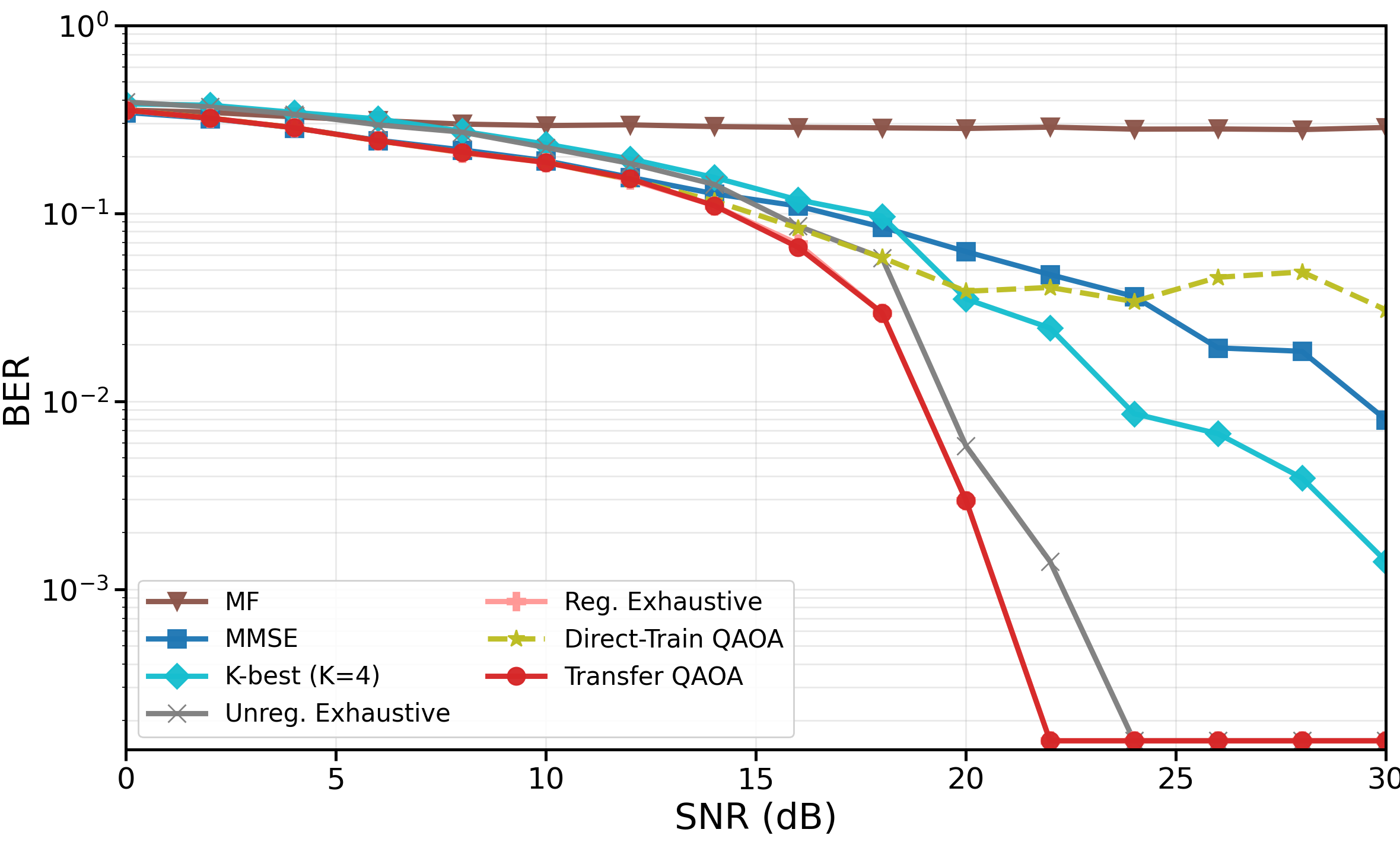}
\caption{Overall BER comparison for the $16\times16$ 16QAM setting. Zero-observed BER points are plotted at the single-error floor for visibility on the logarithmic scale.}
\label{fig:overall_ber}
\end{figure}

Fig.~\ref{fig:overall_ber} compares the BER curves of MF, MMSE, K-best ($K=4$), the regularized blockwise exhaustive detector, the unregularized blockwise exhaustive detector, the direct-training QAOA detector, and the parameter-transfer QAOA detector.

Several trends are important. First, the MF baseline remains far from the other methods across the entire SNR range. Second, the low-SNR region ($0$--$4$ dB) is still dominated by MMSE, whose BER is slightly lower than that of all blockwise variants. Third, once the SNR reaches the medium regime, the regularized blockwise family becomes the strongest group. In particular, the parameter-transfer QAOA detector nearly overlaps the regularized blockwise exhaustive detector over most tested SNR points and clearly improves upon both MMSE and K-best in the medium-to-high-SNR region.

The comparison also reveals a clear distinction between the two QAOA realizations. The direct-training QAOA detector stays close to the transfer curve only at low and moderate SNR, but a noticeable gap appears from about 14 dB onward. By contrast, the parameter-transfer QAOA detector continues to track the regularized exhaustive reference and is the best QAOA-based realization in the average-BER sense, as well as over most of the medium-to-high-SNR range. Averaged over all 16 SNR points, its BER is $1.226\times10^{-1}$, essentially identical to the regularized exhaustive reference ($1.225\times10^{-1}$) and clearly lower than direct-training QAOA ($1.405\times10^{-1}$), MMSE ($1.417\times10^{-1}$), and K-best ($1.606\times10^{-1}$). Therefore, Fig.~\ref{fig:overall_ber} shows that parameter transfer is the preferred QAOA realization within the proposed block-QAOA-aware framework.

\subsection{Effect of the Regularized Blockwise Reference}
\begin{figure}[!t]
\centering
\includegraphics[width=0.98\linewidth]{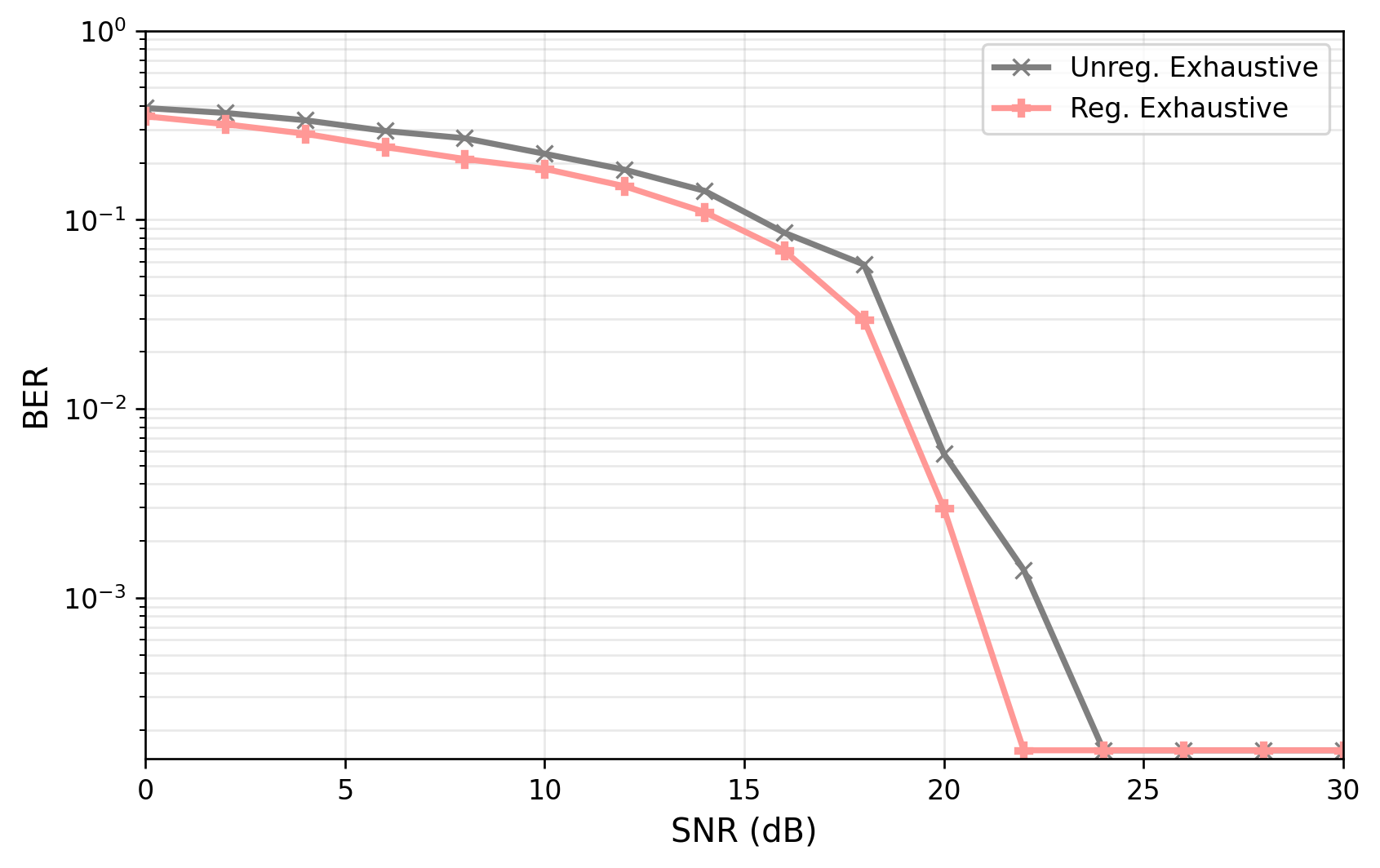}
\caption{BER comparison between the regularized blockwise exhaustive detector and the unregularized blockwise exhaustive detector. Zero-observed BER points are plotted at the single-error floor for visibility on the logarithmic scale.}
\label{fig:ref_ablation_ber}
\end{figure}

Before discussing the QAOA realization, it is necessary to verify that the underlying blockwise objective is itself meaningful. In particular, the present ablation is intended to assess the MMSE-induced dynamic regularization introduced in Section III-C, rather than regularization in an abstract sense. For this purpose, Fig.~\ref{fig:ref_ablation_ber} compares the regularized blockwise exhaustive detector with the unregularized blockwise exhaustive detector obtained by removing the SNR-dependent regularization term $\lambda(\rho)\|\mathbf g_M(\mathbf s_\ell)-\mathbf z_\ell^{\rm MMSE}\|_2^2$ from the local block cost.

The regularized reference consistently improves upon the unregularized exhaustive search over the tested SNR range. For example, at 18 dB the BER of the regularized blockwise exhaustive detector is reduced to $2.953\times10^{-2}$, whereas the unregularized exhaustive counterpart remains at $5.984\times10^{-2}$. A similar gap is observed at 20 dB, where the regularized blockwise exhaustive detector reaches $2.969\times10^{-3}$ while the unregularized exhaustive result is still $6.094\times10^{-3}$. Since the exhaustive solver, block partition, and list widths are otherwise unchanged, this gain can be attributed to the adopted MMSE-induced dynamic regularization schedule itself. In other words, the experiment indicates that the SNR-dependent weight $\lambda(\rho)$ improves the QR-domain blockwise objective in the present detector rather than merely changing the search budget or propagation rule.

This experiment serves as an algorithmic reference study rather than as a main contribution claim. The exhaustive curves are used to validate the adopted dynamic regularized blockwise metric, while the main focus of the paper remains on the QAOA-based realizations built on top of this metric.

\subsection{Effect of Parameter Transfer}
\begin{figure}[!t]
\centering
\includegraphics[width=0.98\linewidth]{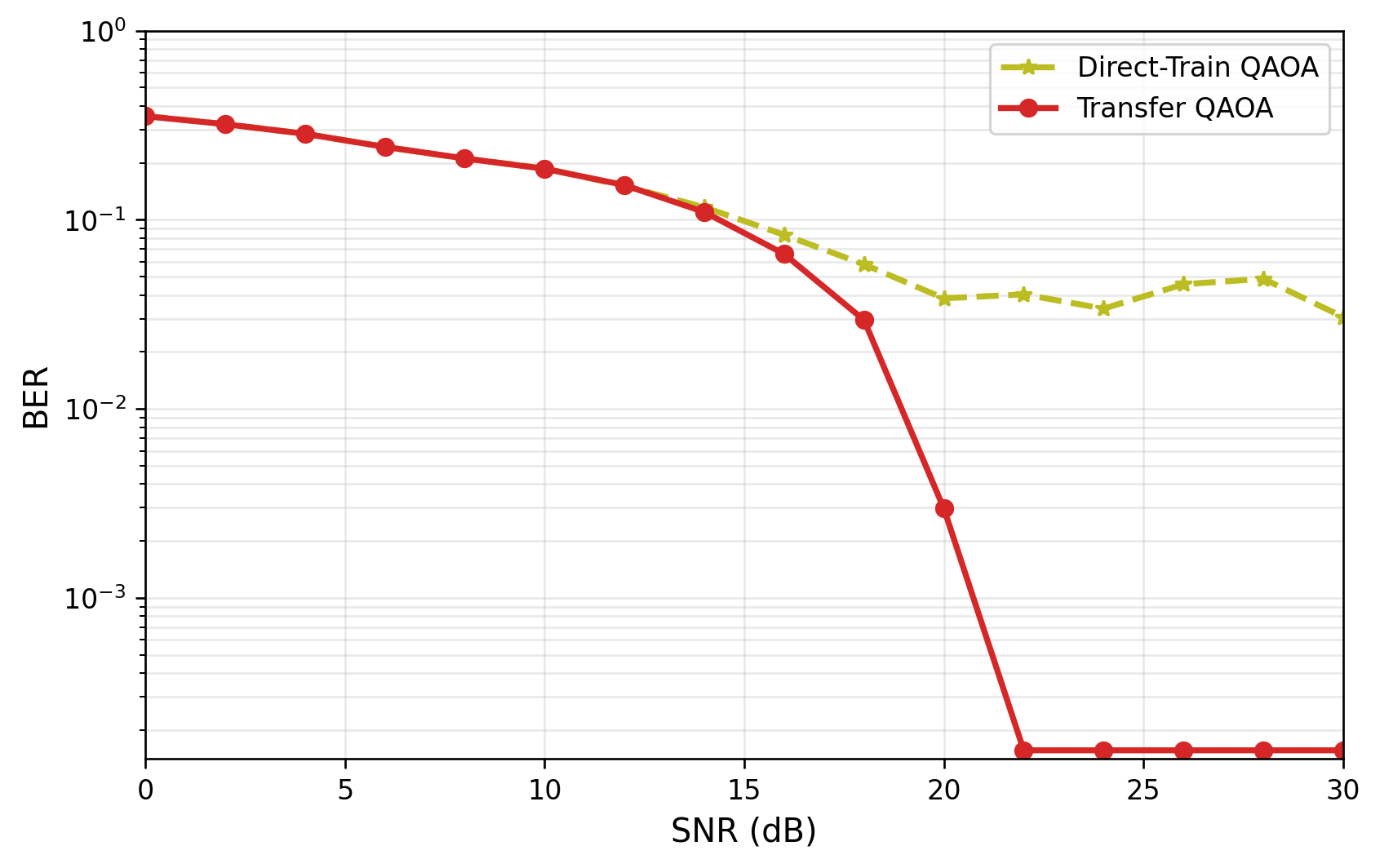}
\caption{BER comparison between direct-training QAOA and parameter-transfer QAOA. Zero-observed BER points are plotted at the single-error floor for visibility on the logarithmic scale.}
\label{fig:transfer_ber}
\end{figure}

Fig.~\ref{fig:transfer_ber} compares the two QAOA-based realizations: the direct-training QAOA detector and the parameter-transfer QAOA detector with block mixing. Parameter transfer clearly outperforms direct training in BER and almost fully recovers the performance of the regularized exhaustive reference.

More specifically, the two QAOA variants are close only in the low-SNR region. At 12 dB their BERs are still similar ($1.527\times10^{-1}$ for transfer and $1.513\times10^{-1}$ for direct training), but from 14 dB onward the gap becomes systematic. At 16 dB the parameter-transfer BER is reduced from $8.328\times10^{-2}$ to $6.594\times10^{-2}$; at 18 dB it drops from $5.813\times10^{-2}$ to $2.953\times10^{-2}$; and at 20 dB it further decreases from $3.844\times10^{-2}$ to $2.969\times10^{-3}$. Over the entire tested range, the average BER of parameter-transfer QAOA is $1.226\times10^{-1}$, compared with $1.405\times10^{-1}$ for direct-training QAOA. These observations indicate that, within the block-QAOA-aware design, fixed-size local subproblems make it possible to reuse structurally matched parameters that are substantially more robust than per-instance online training in the present setting.

From an implementation viewpoint, the transfer realization replaces per-instance local parameter fitting by reuse of an offline template bank together with online selection among structurally matched transferred parameters. In this paper, BER is treated as the primary experimental metric.

\subsection{Discussion of Experimental Scope and Practical Interpretation}
The current experiments support the core conclusions of this paper within the tested $16\times16$ 16QAM setting. First, the overall BER comparison establishes a clear ordering among classical baselines, exact blockwise references, and the two QAOA-based realizations in the target scenario. Second, the comparison between the regularized and unregularized blockwise exhaustive detectors supports the adopted regularized blockwise objective. Third, the direct-training-versus-transfer comparison shows that parameter transfer is the strongest QAOA realization tested in this study and nearly matches the regularized exhaustive reference.

The role of the present experiments should nevertheless be interpreted with appropriate scope. The exhaustive references are included to validate the blockwise metric and provide an algorithmic reference, not to replace the QAOA-based theme of the paper. Moreover, because all QAOA routines are executed on a classical simulator, BER rather than raw wall-clock time is treated as the principal metric in this section. The zero-BER points at high SNR should be interpreted as ``no observed bit errors'' under the present sample size, not as a proof of exact zero error probability. Broader conclusions will require further validation under higher-order modulations, larger antenna dimensions, different block sizes and local qubit budgets, and real quantum hardware or hybrid backends.

\section{Conclusion}
This paper presented BQA-MD, a block-QAOA-aware detector for large-scale MIMO under limited local qubit budget. The central idea is to reorganize the large-scale MIMO detection problem into fixed-size blockwise subproblems whose scale, interface, and path integration are explicitly designed for downstream QAOA-based inference. In this sense, the block-QAOA-aware design is the foundation of the whole framework, while parameter transfer is the key realization strategy enabled by that foundation. Within this framework, the local detector is built on an MMSE-induced blockwise metric with an SNR-dependent dynamic regularization weight.

Simulation results on a $16\times16$ Rayleigh MIMO system with 16QAM showed that the regularized blockwise reference improves upon its unregularized counterpart, validating the adopted blockwise objective and the overall blockwise design rationale. The results also showed that the parameter-transfer QAOA detector nearly matches the regularized exhaustive reference and clearly outperforms direct-training QAOA in BER. This supports parameter transfer as the preferred QAOA realization within the proposed framework: once the large-scale MIMO problem has been reorganized into fixed-size local subproblems, parameter reuse becomes structurally well defined and empirically effective.

The theoretical complexity discussion further clarifies that parameter transfer changes the computational structure by moving parameter-learning effort to offline template-bank construction and eliminating per-instance online variational optimization during detection.

The current implementation is still evaluated on a classical simulator rather than quantum hardware. Future work will include extending the same block-QAOA-aware framework to coded soft-output detection, studying alternative offline transfer-template construction strategies, and investigating hardware-oriented realization under realistic NISQ noise and compilation constraints on practical quantum or hybrid accelerators.

\section*{Acknowledgment}
This work was sponsored by CPS-Yangtze Delta Region Industrial Innovation Center of Quantum and Information Technology-MindSpore Quantum Open Fund.

\end{document}